# Uncovering a factor-based expected return conditioning structure with Regression Trees jointly for many stocks

Vassilis Polimenis[1]

**Abstract**

Given the success and almost universal acceptance of the simple linear regression three-factor model, it is interesting to analyze the informational content of the three factors in explaining stock returns when the analysis is allowed to consider non-linear dependencies between factors and stock returns. In order to better understand factor-based conditioning information with respect to expected stock returns within a regression tree setting, the analysis of stock returns is demonstrated using daily stock return data for 5 major US corporations. The first finding is that in all cases (solo and joint) the most informative factor is always the market excess return factor. Further, three major issues are discussed: a) the balance of a depth=1 tree as it relates to properties of the stock return distribution, b) the mechanism behind depth=1 tree balance in a joint regression tree and c) the dominant stock in a joint regression tree. It is shown that high skew values alone cannot explain the imbalance of the resulting tree split as stocks with pronounced skew may produce balanced tree splits.

## Introduction

The most well known model of market equilibrium is the CAPM (capital asset pricing model) of Sharpe (1964) and Lintner (1965). In this model returns are explained by the slope in the regression of an asset's return on the market return (the market β). CAPM is an intuitively appealing and simple model that has laid the foundations of asset pricing theory, but its major prediction that market β suffices to explain expected returns seems to be violated in many ways (Fama and French, 1992). In modern asset pricing, a model that precludes arbitrage can be summarized by p=Emx with the price of an asset (p) equal to the expected priced cash-flows (x) that the asset can generate, and m a stochastic discount factor. A significant goal is to identify the factors comprising m that best explain expected returns on all assets. An asset pricing model that provides a better description of average returns and has in many ways replaced CAPM is the three-factor model of Fama and French (1993). The three-factor model is based on a time-series regression of excess portfolio returns of the type

$$R(t) - rf(t) = a + b \cdot mex(t) + s \cdot smb(t) + h \cdot hml(t) + e(t) \qquad (1)$$

with R(t) the return on a security or portfolio for period t, rf(t) the risk-free return, mex(t) the excess return on the value-weight (VW) market portfolio above the risk-free asset, smb(t) the return on a diversified portfolio of small stocks minus the return on a diversified portfolio of big stocks, hml(t) the difference between the returns on diversified portfolios of high and low book-to-market (B/M) ratio stocks.

---

[1] polimenis@yahoo.com, Aristotle University of Thessaloniki

If the three-factor sensitivities b, s and h in (1) capture most variation in expected returns, the true value of the intercept a in (1) should be near zero for well -priced securities or portfolios. The success of (1) in explaining returns is such that most empirical research on asset pricing routinely includes the three-factor model among the alternatives. In many instances, when the issue is to measure the performance of a proposed new asset pricing model, an often used benchmark becomes the degree at which the model comes close to explaining as much of the cross-section of average returns as the three-factors.

Given the success and almost universal acceptance of the simple linear regression (1), it is interesting to analyze the informational content of the three factors in explaining stock returns when the analysis is allowed to consider non-linear dependencies between factors and stock returns. One of the significant predictive modelling approaches used in machine learning, and data mining are the Decision trees. Due to their enhanced intelligibility and simplicity Decision trees are among the most popular machine learning algorithms. In particular, Decision trees where the target variable can take continuous values (as in fitting financial return data) are called **regression trees (RT)**.[2] The use of the regression tree for the analysis of single stock returns is demonstrated in Polimenis (2020). In that paper, using daily returns for IBM during the 5-year period from 1/5/2015-30/4/2020 it is found that a) the excess return factor carries the highest informational strength among the three factors and b) when the analysis is limited to SMB and HML, the Small Minus Big return factor is the factor with the most information.

Using the same 5-year period from 1/5/2015-30/4/2020, the analysis here demonstrates joint regression tree analysis for two stocks. The question is not only the informational value of the 3 factors but also how balanced the depth 1 regression tree is and its relation to the underlying stock volatility, skew, and kurtosis. It is found that despite the obvious intuition about a relation of the tree balance to stock return skew, skewness alone is not enough to explain it as there are cases where stocks with high skew produce more balanced trees.

**Methodology**

In the analysis here, stock return data are jointly fit to the 3 Fama-French factors via a Regression Tree (RT), a non-parametric supervised learning method. When using factors with a RT, the goal is to obtain a model that conditions expected US returns by learning simple decision rules inferred from the factors. The analysis is jointly performed for various pairs of major US corporations and for a limited sample period, and thus it is only done for demonstration purposes and not for statistical inference.

An important benefit of using a RT to uncover conditional factor information on returns is that a RT produces a white box model. This means that when a given situation is observable in the model, the conditioning structure is easily understood via boolean logic. By contrast, in a black box model (such as the one produced by an artificial neural network), results can be difficult to interpret. A second

---
[2] For more on Decision Trees see Breiman et. al. (1994), Quinlan (1993) and Hastie, Tibshirani and Friedman (2009)

benefit for the use of Trees is that little data preparation is needed before RTs are employed. On the contrary, other methods require data normalisation, the use of dummy variables etc.

Given n factors stored in tuples X(t) and portfolio returns y(t), a Regression tree recursively partitions the factor space in hyper-rectangles such that the samples with similar portfolio returns are grouped together. Let the sub-sample at node m be represented by Q. For each candidate split $\theta=(j,k_m)$ consisting of a factor j and a threshold value $k_m$ the tree further partitions the sample into the left and right subsets

$$Q_{left}(\theta) = [X(t),y(t)] \mid X_j(t) < k_m$$

$$Q_{right}(\theta) = [X(t),y(t)] \mid Q \setminus Q_{left}(\theta)$$

In order to choose the best split point in a regression tree predictive modeling situation all input variables and all possible split points are evaluated and chosen in a greedy algorithm. The cost function that is minimized when choosing split points is the sum squared error across all training samples against their sub-region prediction

$$Min_{\text{for all split points}} \Sigma_i(y_i - prediction(y_i))^2$$

where y is the target output for the training sample and prediction is the predicted output for the specific rectangle where y is placed. The algorithm maximizes the drop in that value when moving from a node to its children.

Factor data (the features in machine learning terminology) comprise 3 return factors. The first and most well known factor is the excess return on the market[3] above the risk free rate[4]

$$mex(t) = Rm(t) - Rf(t)$$

The other 2 factors are the so-called Fama-French factors. The Fama/French factors[5] are constructed using the 6 value-weight portfolios formed on size and book-to-market.[6] SMB (Small Minus Big) is the average return on the three small portfolios minus the average return on the three big portfolios

SMB = 1/3 (Small Value + Small Neutral + Small Growth) - 1/3 (Big Value + Big Neutral + Big Growth)

HML (High Minus Low) is the average return on the two value portfolios minus the average return on the two growth portfolios,

HML = 1/2 (Small Value + Big Value) - 1/2 (Small Growth + Big Growth)

When using a regression tree with more than one stocks we have a multi-target (or multi-output) tree. If there were no correlation among the n targets, the obvious way to solve the problem would be to build n independent trees, i.e. one for each target. However, stock returns are highly correlated, and thus obtaining one tree per target is not informationally optimal. Instead, when a single model capable of predicting simultaneously n stocks is built correlation information enters the tree structure.

---

[3] Market return is calculated as a value-weight return of all firms incorporated in the US and listed on the NYSE, AMEX, or NASDAQ at the beginning of month t, have good share and price data at the beginning of t, and good return data for period (t).
[4] the one-month Treasury bill rate (from Ibbotson Associates)
[5] Factor data were downloaded from the French data library
https://mba.tuck.dartmouth.edu/pages/faculty/ken.french/index.html
[6] See https://mba.tuck.dartmouth.edu/pages/faculty/ken.french/data_library.html for the description of the 6 size/book-to-market portfolios

This is a first analysis of the following three issues: a) when solo stock analysis is performed, the mechanism behind single depth tree balance b) the mechanism behind single depth tree balance in a joint regression tree and c) the dominant stock in a joint regression tree.

**Results and discussion**

In the analysis here, 5 years of daily stock return data for 5 major US companies: International Business Machines Corp. (IBM), The Coca-Cola Company (KO), The Bank of New York Mellon (BK), The Procter & Gamble Company (PG), and Alphabet Inc (GOOG). Daily data for the period from 1/5/2015 to 30/4/2020 are used. The sample comprises 1259 daily returns. Basic return statistics for the 5 stock returns during the specific period, are shown in Table 1a. In Table 1b the covariance matrix for the 5 stock returns is shown.

A key question that needs to be addressed is the degree of balance of a tree. In an effort to explain the tree balance (or lack of balance thereof) it is natural to measure target return skewness. Skewness for all 5 stock returns for the period is calculated to -0.27, -1.03 , -0.48, 0.585, 0.66 for IBM, KO, BK, PG and GOOG respectively. Skewness and excess kurtosis in a sample plays a significant role in determining the split point. A very positive skewness (as is the case for PG and GOOG) implies that the outliers are on the right, and thus intuitively should tend to push the split to the right of the distribution. Skewness alone is not enough to determine the split. kurtosis (or better excess kurtosis) is also important in the split formation as it determines the magnitude of the tails.

Limiting to max depth = 1, the solo 3-factor regression tree analysis for IBM is shown in Figure 1 (see discussion in Polimenis 2020). The split criterion is based on market excess returns factor.

Next the analysis proceeds with the solo and joint tree regressions of the remaining 4 stocks (see Figures 2, 3, 4, and 5). Joint trees are always done with IBM as it exhibits the lowest skew and kurtosis among the 5 stocks (i.e. the stock closest to normality). The first finding is that in all cases (solo and joint) the most informative factor remains the market excess return factor mex.

With respect to the **KO analysis** (Figure 2), the solo RT differs from that of IBM with the solo split for KO the point mex <= -3.5%. On the left branch, KO returns are expected to have ER = -490bp versus ER=10bp for the right. With respect to the **joint analysis** the split point for the joint regression tree [mex <= - 70bp] is dominated by IBM. On the left branch, IBM returns are expected to have ER = -180bp versus ER=30bp for the right. On the left branch, KO returns are expected to have ER = -110bp versus ER=20bp for the right.

With respect to the **BK analysis** (Figure 3), the split point differs from that of IBM with the solo split for BK the point mex <= -100bp. On the left branch, BK returns are expected to have ER = -260bp versus ER=30bp for the right. With respect to the **joint analysis** the split point for the joint regression tree [mex <= - 90bp] is roughly ⅔ dominated by BK.[7] On the left branch, IBM returns are expected to have ER = -200bp (vs. -180bp for the solo IBM tree) versus ER=20bp on the right. On the left branch, BK returns are expected to have ER = -240bp versus ER=30bp for the right.

---

[7] volatility for BK stock returns at 2.97bp vs. 2.45bp for IBM (see Table 1b)

With respect to the **PG analysis** (Figure 4), the 1st split point for PG differs significantly from that of IBM with the solo split for PG placed well to the right of the distribution at mex <= 300bp. On the left branch, PG returns are expected to have ER = 0 versus ER=500bp for the right. The reason for such an unbalanced tree is interesting and should be further investigated. One approach is to explain it via the highly positive skew for PG returns at .585. Yet highly positive skew alone is not enough to explain it as GOOG (see Figure 5) has even more pronounced skew at .66 but produces a more balanced tree split. With respect to the **joint analysis** (Figure 4), we observe that the split point for the joint regression tree [mex <= - 70bp] is again (similarly with the other highly skewed tree case of KO) dominated by IBM. On the left branch, PG returns are expected to have ER = -100bp versus ER=20bp for the right.

With respect to the **GOOG analysis** (Figure 5) and despite GOOG returns obtaining the highest skew at .66, the RT appears quite balanced at mex <= -40bp. On the left branch, GOOG returns are expected to have ER = -140bp versus ER=50bp for the right. With respect to the **joint analysis** (Figure 5), we observe that the split point for the joint regression tree [mex <= - 70bp] is again dominated by the IBM split despite the fact that GOOG is more volatile at 2.92bp (vs 2.45bp for IBM), more skewed at .66 (vs -.27 for IBM), has more kurtosis at 11.90 (vs 10.90 for IBM) and attains a more balanced depth 1 tree at 22-78% (vs 13.66 - 86.34% for IBM). On the left branch, GOOG returns are expected to have ER = -190bp versus ER=40bp for the right.

Thus, stock return skewness does not seem to explain tree imbalance as the most balanced tree is attained by GOOG that also happens to have the highest positive skew of .66 in the sample. On the contrary KO with the most negative return skew (-1.07) produces an unbalanced tree. A second observation is that in many cases, IBM that has the least skew and kurtosis in the set of the 5 stocks, seems to dominate depth=1 joint trees (for KO, PG and GOOG). Joint tree between IBM and BK is ⅔ dominated by BK.

**Conclusion**

In order to better understand factor-based conditioning information with respect to expected stock returns, and given the almost universal acceptance of the simple linear regression three-factor model, the use of an important machine learning technique, the regression tree for the joint analysis of stock returns demonstrated. Daily stock return data of 5 major US corporations (IBM, KO, BK, PG and GOOG) are fit to the 3 factors (market excess return, SMB and HML) via a Regression Tree (RT). The strongest finding is that in all cases (solo and joint) the most informative factor is the market excess return factor. Three major issues are discussed: a) when solo stock analysis is performed, the balance of a depth=1 tree as it relates to properties of the stock return distribution, b) the mechanism behind depth=1 tree balance in a joint regression tree and c) the dominant stock in a joint regression tree. The explanation behind an unbalanced tree is interesting and should be further investigated. Despite an obvious intuition to attribute tree imbalance to return skew, it is shown that high skew values alone cannot explain the phenomenon as stocks with pronounced skew may produce balanced tree splits.

**References**

Breiman L., J. Friedman, R. Olshen, and C. Stone. (1994) Classification and Regression Trees. Wadsworth, Belmont, CA

Fama, Eugene F., and Kenneth R. French. (1992), The Cross-Section of Expected Stock Returns. The Journal of Finance, 47: 427-465.

Fama, Eugene F., and Kenneth R. French. (1993) "Common Risk Factors in the Returns on Stocks and Bonds." Journal of Financial Economics 33:3–56

Hastie T., R. Tibshirani and J. Friedman. (2009) Elements of Statistical Learning, Springer

Lintner, John (1965) "The Valuation of Risk Assets and the Selection of Risky Investments in Stock Portfolios and Capital Budgets." Review of Economics and Statistics 47:13–37.

Polimenis, V. (2020) "Uncovering factor-based expected return conditioning structure with Regression Trees for a stock" working paper

Quinlan J.R. (1993) C4. 5: programs for machine learning. Morgan Kaufmann

Sharpe, William F. (1964) "Capital Asset Prices: A Theory of Market Equilibrium Under Conditions of Risk." Journal of Finance 19:425–42

**Table 1a.** Basic daily return statistics for 5 major US companies: International Business Machines Corp. (IBM), The Coca-Cola Company (KO), The Bank of New York Mellon (BK), The Procter & Gamble Company (PG), and Alphabet Inc (GOOG). Daily data for the period from 1/5/2015 to 30/4/2020 are used. The sample comprises 1259 daily returns.

|       | ibm         | ko          | bk          | pg          | goog        |
|-------|-------------|-------------|-------------|-------------|-------------|
| count | 1259.000000 | 1259.000000 | 1259.000000 | 1259.000000 | 1259.000000 |
| mean  | 0.000050    | 0.000298    | 0.000136    | 0.000516    | 0.000876    |
| std   | 0.015640    | 0.011788    | 0.017223    | 0.012666    | 0.017086    |
| min   | -0.128507   | -0.096725   | -0.145034   | -0.087373   | -0.111008   |
| 25%   | -0.006146   | -0.004122   | -0.006570   | -0.004516   | -0.005847   |
| 50%   | 0.000512    | 0.000479    | 0.000637    | 0.000456    | 0.000792    |
| 75%   | 0.006672    | 0.005449    | 0.008348    | 0.005827    | 0.008548    |
| max   | 0.113011    | 0.064796    | 0.156165    | 0.120090    | 0.160524    |

**Table 1b.** Variance - covariance matrix for daily returns for 5 major US companies: International Business Machines Corp. (IBM), The Coca-Cola Company (KO), The Bank of New York Mellon (BK), The Procter & Gamble Company (PG), and Alphabet Inc (GOOG). Daily data for the period from 1/5/2015 to 30/4/2020 are used.

|      | ibm      | ko       | bk       | pg       | goog     |
|------|----------|----------|----------|----------|----------|
| ibm  | 0.000245 | 0.000095 | 0.000150 | 0.000098 | 0.000141 |
| ko   | 0.000095 | 0.000139 | 0.000089 | 0.000097 | 0.000083 |
| bk   | 0.000150 | 0.000089 | 0.000297 | 0.000090 | 0.000146 |
| pg   | 0.000098 | 0.000097 | 0.000090 | 0.000160 | 0.000087 |
| goog | 0.000141 | 0.000083 | 0.000146 | 0.000087 | 0.000292 |

**Table 2a.** Basic daily factor return statistics for the 3 factors during the 5-year period 1/5/2015 to 30/4/2020

|       | mex       | smb       | hml       |
|-------|-----------|-----------|-----------|
| count | 1259.000000 | 1259.000000 | 1259.000000 |
| mean  | 0.000367  | -0.000117 | -0.000356 |
| std   | 0.012019  | 0.005656  | 0.006659  |
| min   | -0.120000 | -0.036200 | -0.047200 |
| 25%   | -0.003250 | -0.003500 | -0.003950 |
| 50%   | 0.000500  | -0.000200 | -0.000600 |
| 75%   | 0.005200  | 0.003100  | 0.002900  |
| max   | 0.093400  | 0.056000  | 0.031400  |

**Table 2b.** Skewness and kurtosis for the 3 factors

|          | mex   | smb  | hml   |
|----------|-------|------|-------|
| skew     | -.73  | .64  | -.12  |
| kurtosis | 20.55 | 9.85 | 6.85  |

**Table 2c.** Various solo (for IBM, KO, BK, PG and GOOG) and joint (one of KO, BK, PG and GOOG jointly with IBM) depth =1 trees fit based on the 3-factor model.

|  | IBM | KO | BK | PG | GOOG |
|---|---|---|---|---|---|
| variance | 2.45bp | 1.39bp | 2.97bp | 1.60bp | 2.92bp |
| skew | -.27 | -1.03 | -.48 | .585 | .66 |
| kurtosis | 10.90 | 13.50 | 16.50 | 16.00 | 11.90 |
| 1st split | -70bp | -350bp | -100bp | 300bp | -40bp |
| Solo tree balance | 13.66 - 86.34% | 1-99% | 9.5-90.5% | 99-1% | 22-78% |
| Joint tree balance | na | 13.66 - 86.34% | 10.5-89.5% | 13.66 - 86.34% | 13.66 - 86.34% |
| 1st Joint split | na | -70bp | -90bp | -70bp | -70bp |

**Figure 1.** Single-target [i.e. separate] 3-factor IBM analysis plot with depth = 1. We observe that a) the most informative factor is X[0] (i.e. the market excess return factor mex), and b) the split point is mex <= - .7%. On the left branch, IBM returns are expected to have ER = -180bp versus ER=30bp for the right.

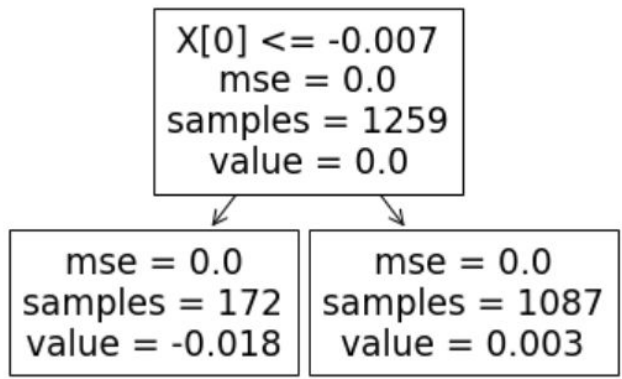

**Figure 2.** Single-target [i.e. separate] 3-factor KO and Multi-target [i.e. combined] 3-factor joint IBM and KO analysis plots with depth = 1.

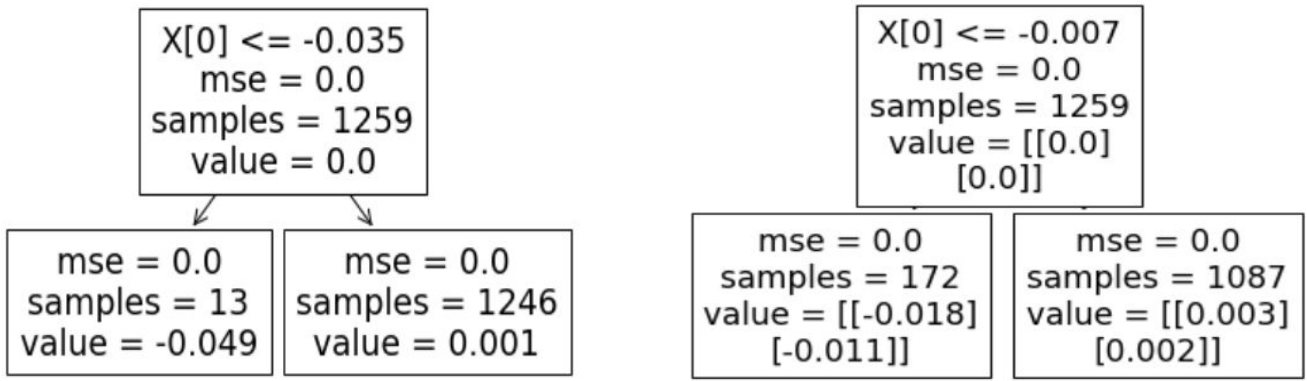

**Figure 3.** Single-target [i.e. separate] 3-factor BK and Multi-target [i.e. combined] 3-factor joint IBM and BK analysis plots with depth = 1.

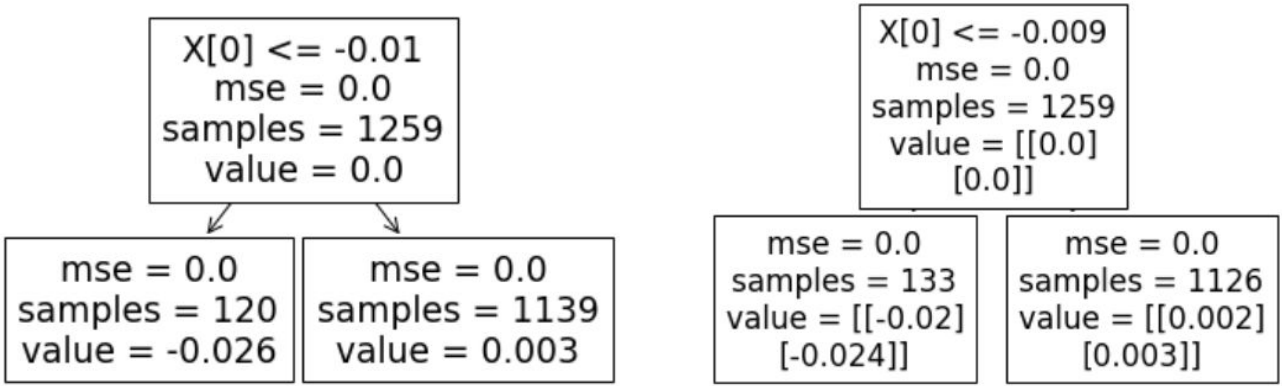

**Figure 4.** Single-target [i.e. separate] 3-factor PG and Multi-target [i.e. combined] 3-factor joint IBM and PG analysis plots with depth = 1.

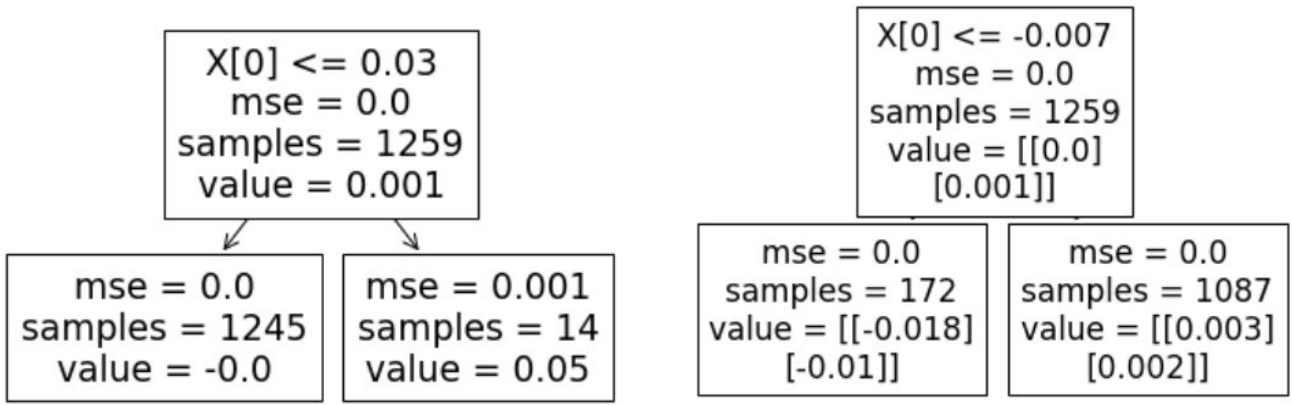

**Figure 5.** Single-target [i.e. separate] 3-factor GOOG and Multi-target [i.e. combined] 3-factor joint IBM and GOOG analysis plots with depth = 1.

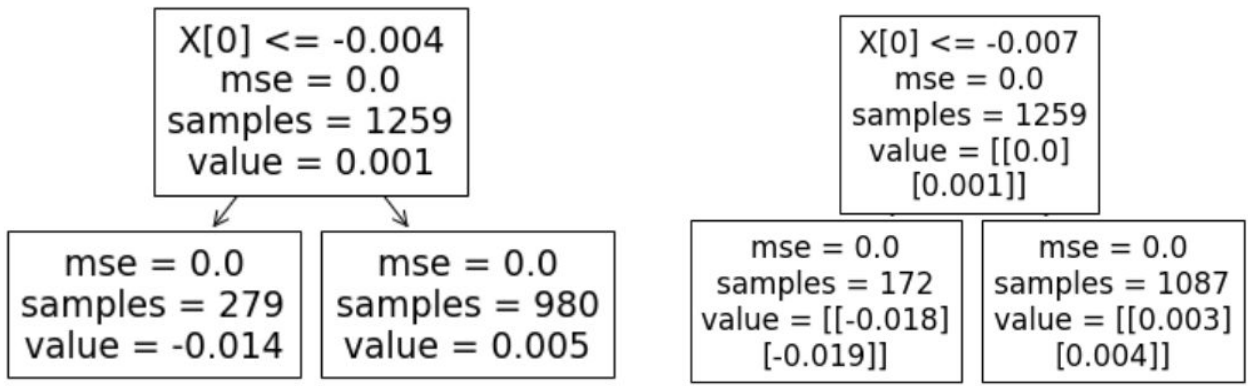